\documentclass[fleqn,12pt,twoside]{article}
\usepackage[headings]{espcrc1}
\usepackage{amsmath}

\readRCS
$Id: espcrc1.tex,v 1.2 2004/02/24 11:22:11 spepping Exp $
\usepackage{graphicx}
\usepackage[figuresright]{rotating}

\newcommand{\be}{\begin{equation}}
\newcommand{\ee}{\end{equation}}

\title{Equation of state from lattice QCD}

\author{S. D. Katz\address[MCSD]{Institute for Theoretical Physics, 
        E\"otv\"os University, Budapest\\
        P\'azm\'any P. s\'et\'any 1/A, H-1117 Budapest, Hungary}}

\runtitle{Equation of state from lattice QCD}
\runauthor{S. D. Katz}

\begin{document}

\maketitle

\begin{abstract}
Recent results on the equation of state from lattice QCD are reviewed.
The lattice technique and previous results are shortly discussed. New 
results for physical quark masses and two sets of lattice spacings are
presented. The pressure, energy density, entropy density, speed of sound 
and quark number susceptibilities are determined.
\end{abstract}

\section{Introduction}
The equation of state (EoS) of strongly interacting matter
plays an important role in understanding heavy ion collisions.
It is an important input for different hydrodynamical models.
In QCD with increasing temperatures ($T$) we expect a transition 
at some $T=T_c$. The dominant degrees of freedom are hadrons in the low 
temperature phase and colored objects in the high temperature
phase. Whether there is a real first order phase transition or only a rapid
cross-over is still an open question.

Since we are mostly interested in the EoS around $T_c$, non-perturbative
methods are necessary among which lattice QCD is the most systematic one. 
There are at least two serious 
difficulties with lattice simulations. The first
one is connected to the lightness of the quark masses. The cost of 
computations increases strongly as the quark masses decrease, therefore
most lattice results were obtained with unphysically large quark masses.
The second difficulty is connected to the continuum limit. Calculations are
always performed at a finite lattice spacing ($a$). In order to
get physical results, we have to take the $a \rightarrow 0$ limit. Since
for the EoS the computational costs scale as $a^{-13}$ it is not surprising
that up to very recently most results were obtained only at one set of 
lattice spacings.

The situation is much easier in the case of the pure gauge theory.
The first problem does not exist since the quark masses are infinite.
There are continuum extrapolated results both with unimproved and improved
lattice actions and they show nice 
agreement~\cite{hep-lat/9602007,hep-lat/9905005,hep-lat/0105012}.
There are also numerous results for the full theory with dynamical quarks
which will be discussed in the following.


\section{Lattice formulation}
Thermodynamical quantities can be obtained from the partition function
which can be given by a Euclidean path-integral:
\be
Z=\int dU d\bar{\Psi} d\Psi e^{-S_E(U,\bar{\Psi},\Psi)},
\ee
where $U$ and $\bar{\Psi},\Psi$ are the gauge and fermionic fields and
$S_E$ is the Euclidean action. The lattice regularization of this 
action is not unique. There are several possibilities to use improved actions
which have the same continuum limit as the straightforward unimproved ones. 
The advantage of improved actions is that the discretization errors are 
reduced and therefore a reliable continuum extrapolation is possible already 
from larger lattice spacings. On the other hand, calculations with improved 
actions are usually more expensive than with the unimproved one.

Usually $S_E$ can be split up as $S_E=S_g+S_f$ where $S_g$ is the
gauge action containing only the self interactions of the gauge fields and
$S_f$ is the fermionic part. The gauge action has one parameter, the 
$g$ gauge
coupling, while the parameters of $S_f$ are the $m_q$ quark masses 
(and in finite density studies the chemical potentials). For the fermionic
action the two most widely used discretization types are the Wilson and
staggered fermions.

For the actual calculations finite lattice sizes of $N_s^3N_t$ are
used. The physical volume and the temperature are related to the lattice
extensions as:
\begin{align}
V=(N_sa)^3, && T=\frac{1}{N_ta}.
\end{align}  
Therefore lattices with $N_t \gg N_s$ are referred to as zero temperature
lattices while the ones with $N_t<N_s$ are finite temperature lattices.

For large homogeneous systems the pressure is proportional to the free
energy density. Unfortunately the free energy density ($-T/V\log Z$) cannot be measured 
directly. We can only measure the derivatives of $\log Z$ with respect
to the parameters of the action. Then, with an integration we can 
obtain the pressure. This method is known as the integral method for
calculating the pressure. In order to remove the divergent zero-point energy
we have to subtract the pressure measured on zero temperature lattices.
This subtraction makes the determination of the equation of state
so difficult, since small statistical errors are needed to get reasonable 
uncertainties after the subtraction.
Further thermodynamical quantities can be derived directly from the pressure.
For example the energy density ($\epsilon$), entropy density ($s$) and
speed of sound ($c_s$) have the following relation with the pressure:
\begin{align}
\label{eq:escs}
\epsilon = T(\partial p/\partial T)-p, && s = (\epsilon + p) T, && c_s^2
=\frac{dp}{d\epsilon}.
\end{align}

There is another way to obtain the energy density, which is often used in 
the literature. One can calculate the quantity $\epsilon-3p$ directly
from the partition function. 

\begin{figure}
\centerline{
\includegraphics*[width=6.3cm]{eos_milc.eps}\hspace*{1cm}
\includegraphics*[width=9cm]{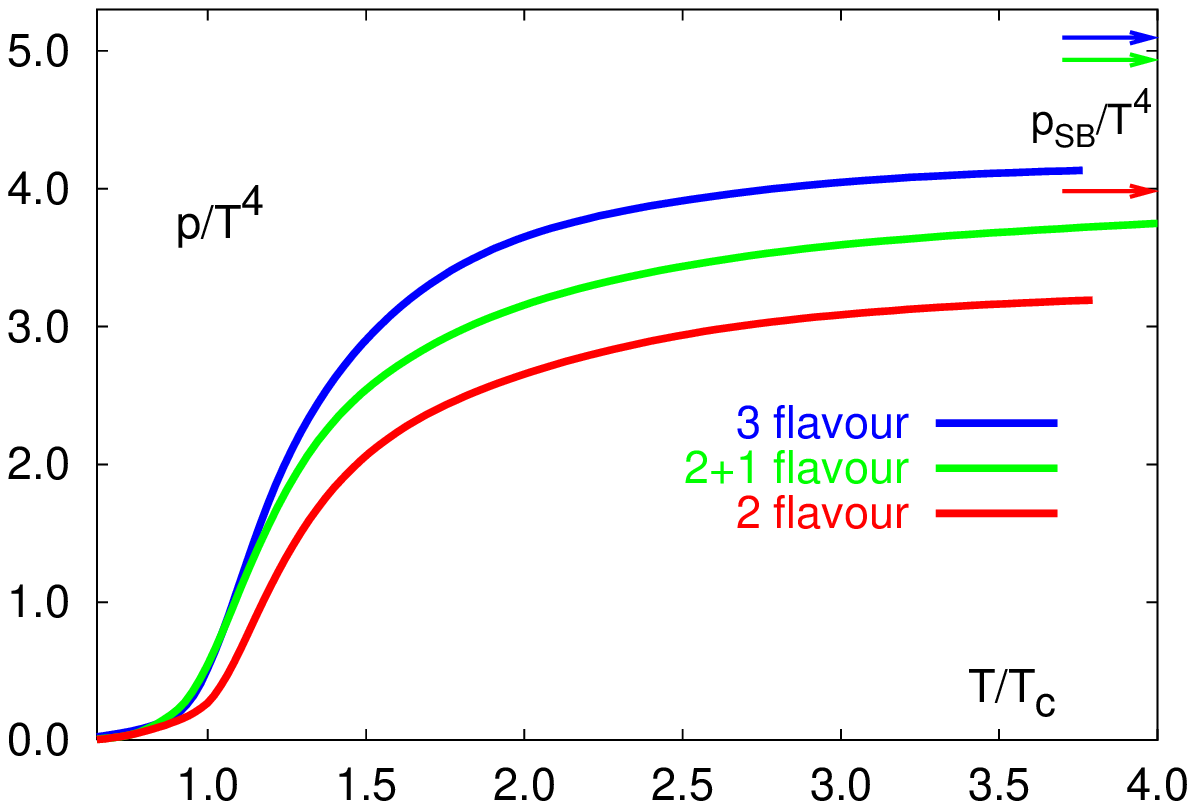}}
\caption{{\em Left:} The pressure (lower symbols) and energy
density (upper symbols) with 2 flavors of unimproved staggered 
fermions on $N_t=4$ (diamonds) and $N_t=6$ (squares and circles) 
lattices for different $m_q/T$ ratios.~\cite{hep-lat/9612025}.
{\em Right:} The pressure 2, 2+1 and 3 flavors of $p4$ improved
staggered fermions on $N_t=4$ lattices~\cite{hep-lat/0002003}.
}
\label{Staggered}
\end{figure}

\begin{figure}
\centerline{\includegraphics*[width=8.5cm]{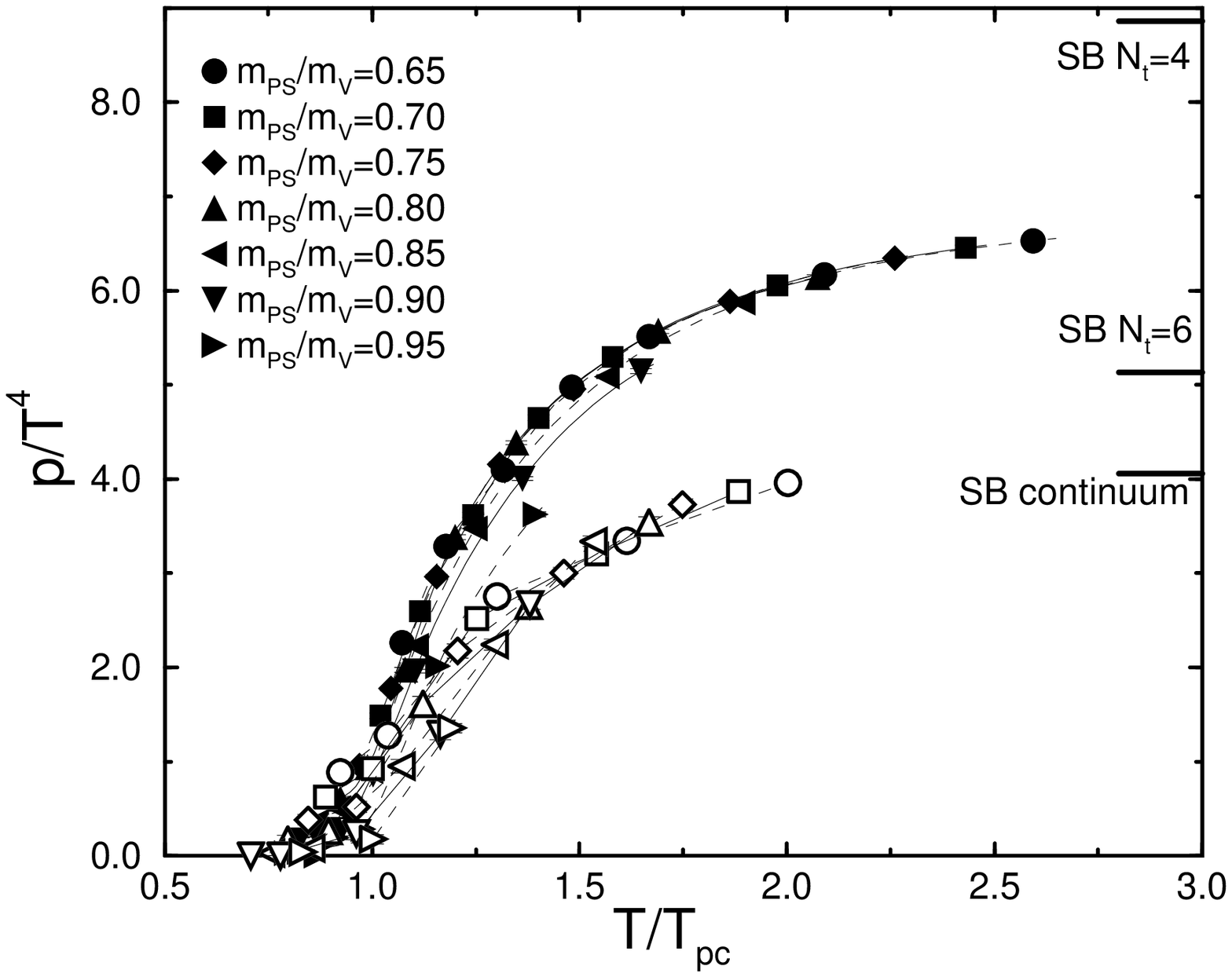}
\includegraphics*[width=8.5cm]{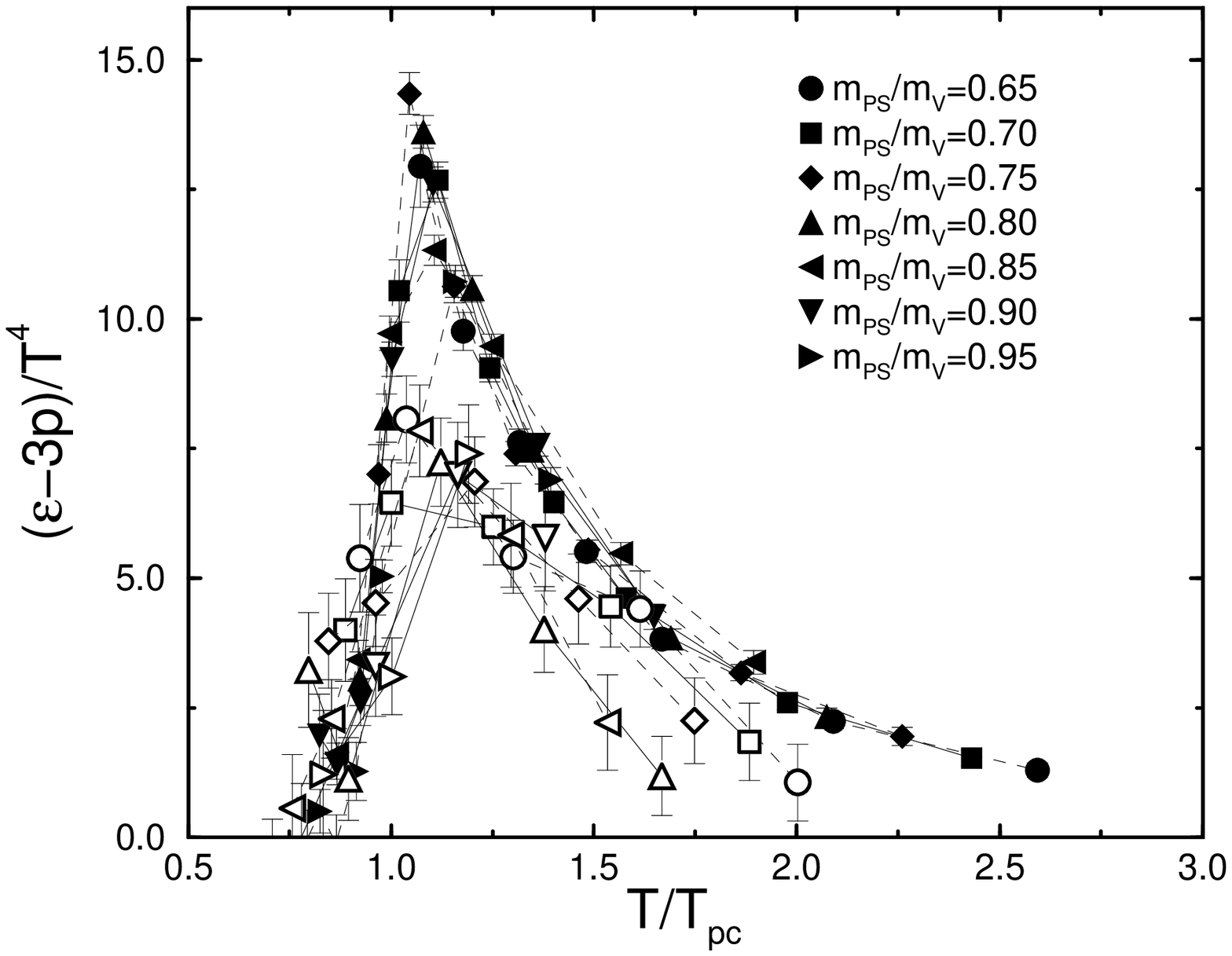}}
\caption{The pressure (left) and $\epsilon-3p$ (right) obtained from
{\cal O}(a) improved Wilson fermions for several pion 
masses on $N_t=4$ (filled symbols) and $N_t=6$ (open symbols) 
lattices~\cite{hep-lat/0103028}}
\label{Wilson}
\end{figure}

\section{Line of constant physics}
Lattice calculations of the EoS are usually performed with a fixed $N_t$
and then, since in a fixed temperature range $N_t$ 
is inversely proportional to the lattice spacing,
the continuum limit can be approached by increasing $N_t$.
Keeping $N_t$ constant  means that the temperature can only be varied by changing
the lattice spacing. This is usually achieved by varying the $g$ 
gauge coupling.
If we want to keep e.g. the quark masses constant
then the dimensionless lattice mass parameters ($am_q$) have to be tuned accordingly.
This defines the line of constant physics (LCP) in the parameter space.

If we keep the mass parameters constant and do not follow the LCP
-- which is the case in most EoS 
lattice studies -- then we have to face the following unphysical situation.
Cooling down two systems, one at $3T_c$ and one at $T_c$ to zero temperature,
the quarks in the former case will be 3 times heavier. In this
approach not $m_q$ but $m_q/T$ is kept constant.

The determination of the LCP is expensive since it requires hadron
spectrum measurements at several lattice spacings. Furthermore, following
the LCP during the determination of the EoS makes the computations
more difficult, since we have to decrease the quark mass parameters 
with increasing temperatures.
These are the reasons why most previous works ignored this physically 
important step of the analysis.

\section{Previous results}
There are numerous lattice results for the EoS using dynamical quarks.
However, in all cases the quark masses
-- for computational reasons mentioned in the introduction -- were set
to higher values than their physical one. The first results were obtained
with staggered fermions. Calculations were performed by the MILC 
collaboration~\cite{hep-lat/9410014,hep-lat/9612025} 
and by Karsch, Laermann and Peikert 
from Bielefeld~\cite{hep-lat/0002003}. The first 
calculation with Wilson fermions was done
by the CP-PACS collaboration~\cite{hep-lat/0103028}.

Staggered results are shown on Fig. \ref{Staggered}. 
No LCP was used in these cases, which means that the curves
correspond to constant $m_q/T$, i.e. increasing quark masses with increasing
temperature. 
Fig. \ref{Wilson} shows the EoS obtained with Wilson fermions for $N_t=4$ 
and 6. The lowest quark mass used here corresponds to a pion mass of 
$\approx 500$~MeV. The LCP was used in this analysis.

In the last years small nonzero chemical
potentials~\cite{Fodor:2001au,Fodor:2001pe} 
have also been used to determine the EoS~\cite{hep-lat/0208078,hep-lat/0303013,hep-lat/0305007,hep-lat/0401016}.

Recently, at the lattice conference 
both the MILC collaboration~\cite{hep-lat/0509053} and the
RBC-Bielefeld collaboration~\cite{Jung:2005yb} reported on their ongoing work
in QCD thermodynamics.

Although the published results all apply QCD with dynamical quarks they still
have several weaknesses.
\begin{enumerate}
\item In all cases, unphysical quark masses were used, which results
in unphysical pion masses. Since the transition temperature is higher 
than the physical mass of the pion, but smaller than the pion masses
used in these calculations, it might be important to use physical values.

\item The works with staggered fermions did not use the line
of constant physics which results in an unphysical dependence
of the hadron masses on the temperature.

\item A known problem with staggered fermions is the taste
symmetry violation which causes a non-physical non-degeneracy of
the pion masses. This non-degeneracy dissapears in the continuum
limit, but it is still large for the lattics spacings used in
these calculations.

\item The approximate R algorithm~\cite{Gottlieb:1987mq} was used for the calculations
with 2,3 or 2+1 flavors of staggered fermions. This
algorithm has an intrinsic stepsize which leads to systematic errors
in the results. In order to eliminate this systematics an extrapolation
to zero stepsize should be performed. None of the previous works have
done such an extrapolation. It should be mentioned that due to the 
subtraction in the calculation of the pressure the error coming
from the typically used finite stepsizes is comparable with the
result itself.

\item The discretization errors are still probably large. This is especially
true for temperatures around and below $T_c$ where the lattice spacing of
$N_t=4$ lattices can be as large as 0.3~fm.

\item The determination of the physical scale is not always
unambiguous. Ref.~\cite{hep-lat/0002003} uses e.g. the string tension 
which is -- strictly speaking -- not an existing quantity in full QCD
since at large distances the string breaks and a meson pair is produced.
\end{enumerate}

\section{New results with physical quark masses}
In the following the new results obtained in collaboration with
Y. Aoki, Z. Fodor and K.K. Szab\'o are presented. Details of
this work are found in Ref.~\cite{AFKSZ}.
We have determined the EoS for two sets of lattice spacings, $N_t=4$ 
and $6$. We improved on all points listed above.

\subsection{Lattice action, LCP}
\begin{figure}
\begin{center}
\includegraphics*[height=7.3cm]{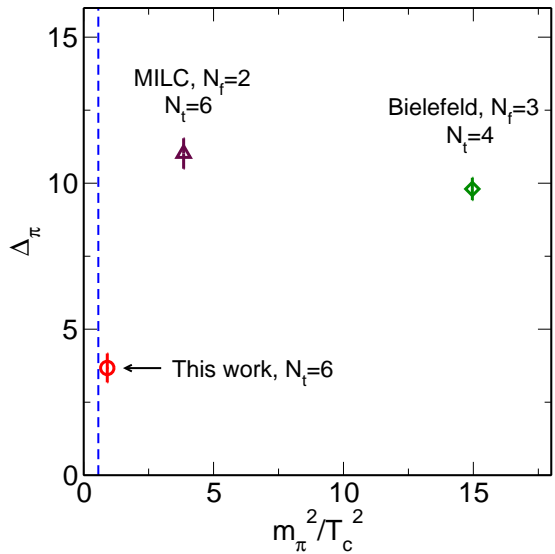}
\end{center}
\caption{
 Pion mass splitting $\Delta_\pi=(m_\pi'^2-m_\pi^2)/T_c^2$ as a 
 function of $(m_\pi/T_c)^2$. The lattice spacings are the same as those at the
 finite temperature transition point. 
The mass of the Goldstone pion is denoted by $m_\pi$, that of the first
 non-Goldstone mode is by $m_\pi'$. The horizontal blue line corresponds to the physical value of
$(m_\pi/T_c)^2$, where $T_c= 173$MeV was assumed \cite{Karsch:2000kv}.
}
\label{taste}
\end{figure}

The lattice action we used was a combination of the
tree level Symanzik-improved gauge action 
and the stout-improved fermionic action~\cite{hep-lat/0311018}. 
The stout improvement is known to reduce 
the taste symmetry violation significantly. 
This is illustrated on Fig. \ref{taste}
where the pion mass splittings of the stout-improved action, 
the Bielefeld p4 action 
and the unimproved staggered 
action (used by previous MILC calculations) are compared. 

As mentioned above, using an approximate algorithm without performing
the necessary extrapolations is dangerous. Instead we used the exact
rational hybrid Monte-Carlo (RHMC) 
algorithm~\cite{hep-lat/0209035,hep-lat/0309084}.

The quark masses were set to their 
physical values so that the meson masses agree with their physical values
up to a few percent. Moreover, the physical 
quark masses were kept constant while increasing the temperature, i.e. we
followed the LCP. In order to determine the LCP we performed three flavor 
$T=0$
simulations. According to leading order chiral perturbation theory, if
we set the masses of all three flavors to the mass of the strange quark, 
we have
\begin{equation}
 m_{PS}^2/m_{V}^2|_{m_q=m_s} = (2m_K^2-m_\pi^2)/m_\phi^2,
  \label{eq:tl chpt}
\end{equation}
This equation was used to set the strange quark mass for different 
lattice spacings (different temperatures at finite $T$). For the light
quarks we used $m_{ud}=m_s/25$. This approximate determination of the LCP
turned out to be sufficient. The uncertainties coming from the LCP were
negligible.

As discussed previously, the determination of the EoS requires both
$T>0$ and $T=0$ simulations. In the finite temperature case the
derivatives of the partition function were determined for several gauge
couplings (we had 16 points for $N_t=4$ and 14 points for $N_t=6$) and
then the integral method was applied to get the pressure. 
At $T=0$ chiral perturbation theory can be used to give 
the quark mass dependence of the chiral condensates needed for the integration.
Therefore we used four pion masses -- somewhat larger then the 
physical one -- and performed the integration for the pressure with the
help of chiral perturbation theory. The four pion masses corresponded
to 3,5,7 and 9 times the physical quark masses.

In order to give the EoS in $T/T_c$ units, we had to find the ratio of the
scales at the different simulation points. For this we matched the
static quark-antiquark potential for the different points 
at an intermediate distance. $T_c$ was defined as the turning point of the
isospin number susceptibility ($\chi_{I}$, see later). The precise 
determination of $T_c$, i.e. connecting the scale to physical quantities
will be the subject of a subsequent publication.

\subsection{Results}
In order to present the $N_t=4$ and $6$ results on the same plots 
we rescale all quantities in the following way. At infinite temperatures
all quantities should approach their free Stefan-Boltzmann limit ($c$).
This limit is, however different in the continuum ($c_{cont}$) and on 
lattices with some fixed $N_t$ ($c_{N_t}$). Therefore all results are 
scaled with a factor $c_{cont}/c_{N_t}$ so that they could be compared
with the continuum  Stefan-Boltzmann limits.

\begin{figure}
\begin{center}
\includegraphics*[width=7.9cm]{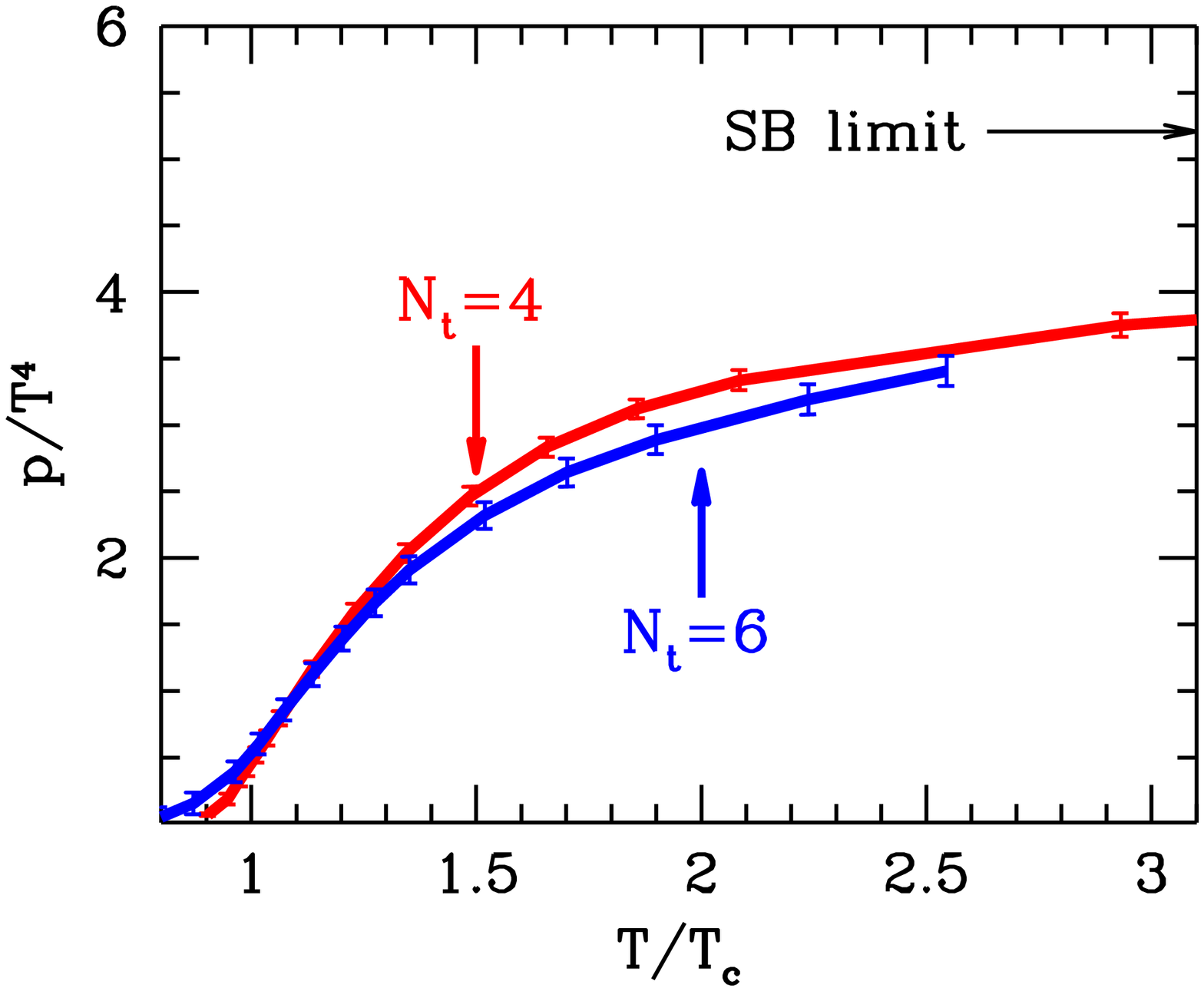}
\includegraphics*[width=7.9cm]{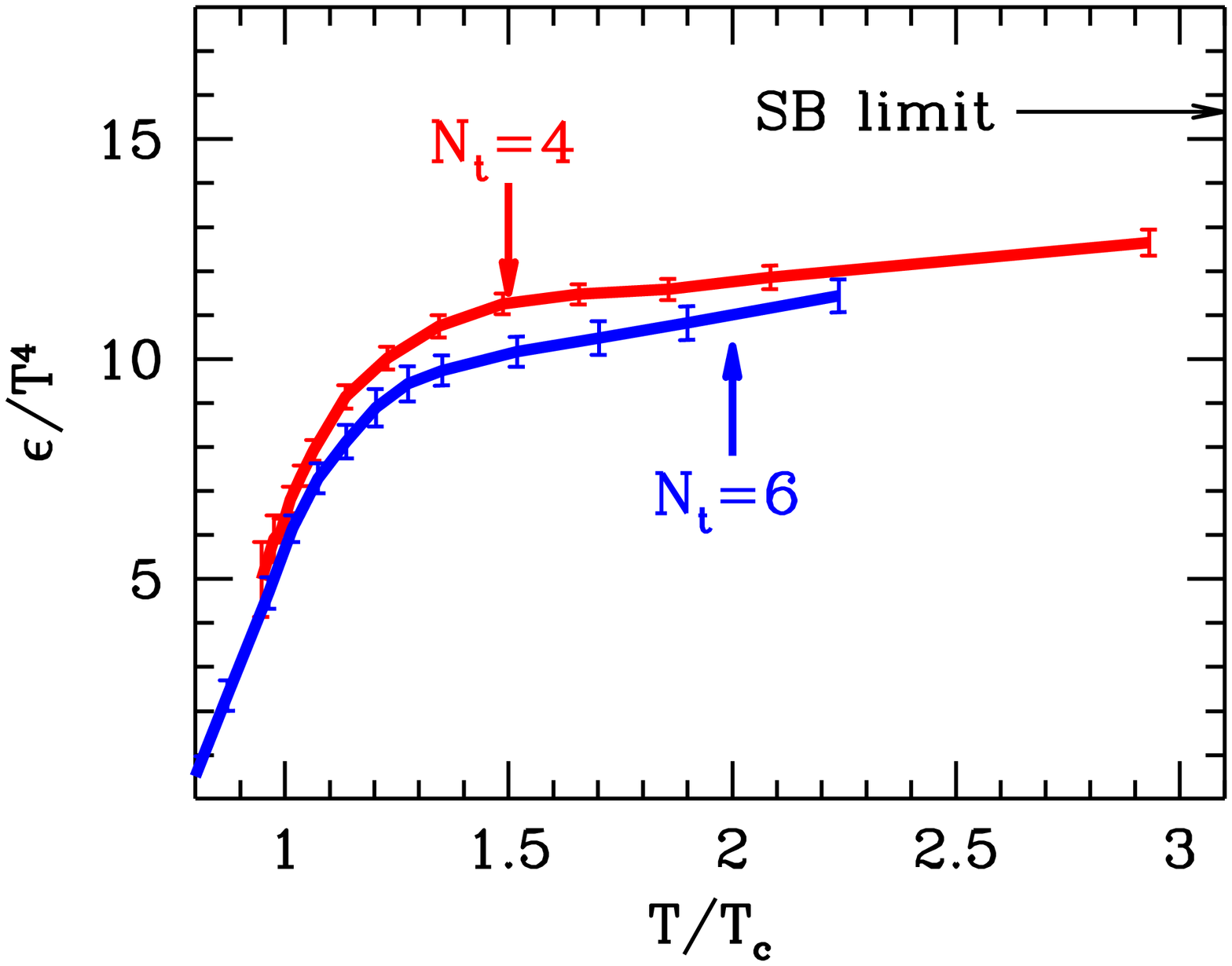}
\end{center}
\caption{\label{eos_pe}
{\em Left:} the pressure $p$,
as a function of the temperature. Both $N_t$=4 (red, upper curve)
and $N_t$=6 (blue, lower curve) data are obtained along the LCP. They 
are normalized by $T^4$ and scaled by $c_{cont}/c_{N_t}$. 
In order to lead the eye lines connect the data points.
{\em Right:} the energy density ($\epsilon$), red (upper) and blue (lower) 
for $N_t$=4 and 6 respectively. 
This result was obtained directly from the pressure. 
}
\end{figure}

\begin{figure}
\begin{center}
\includegraphics*[width=7.9cm]{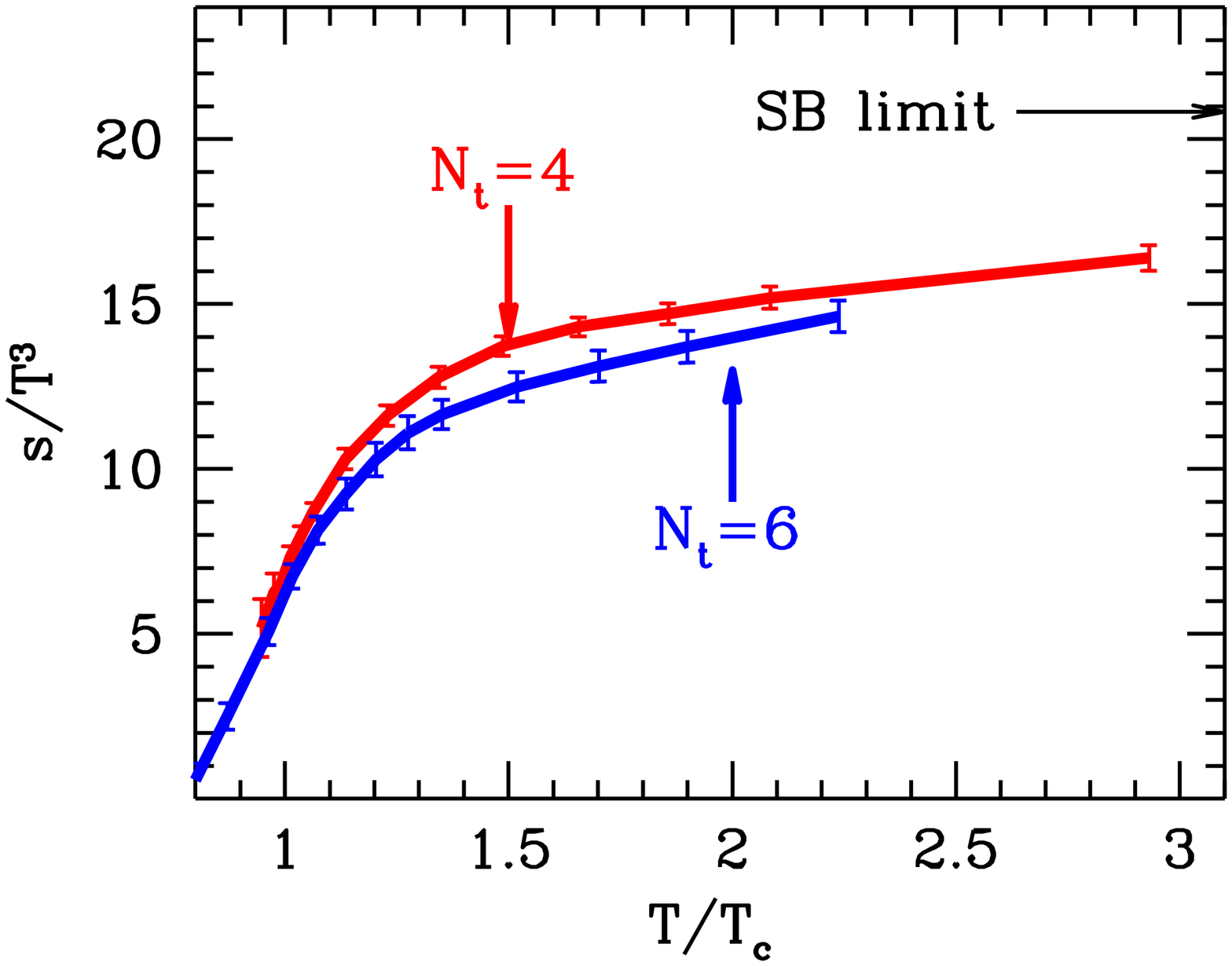}
\includegraphics*[width=7.9cm]{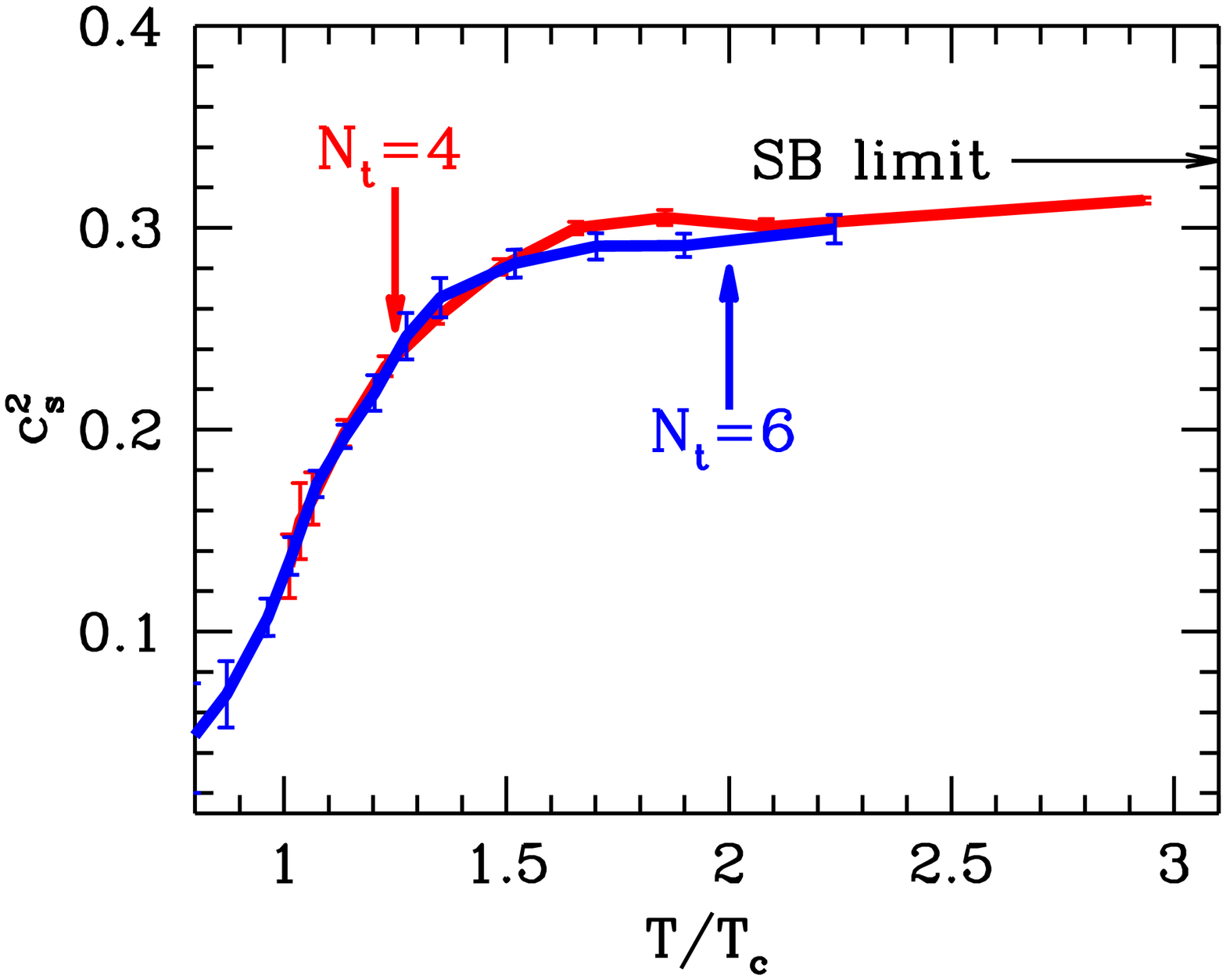}
\end{center}
\caption{\label{eos_sc}
{\em Left:} the 
entropy density, normalized by $T^3$. 
{\em Right:} the speed of sound squared. 
The labeling is the same as for Figure \ref{eos_pe}.
}
\end{figure}

Fig. \ref{eos_pe} shows the pressure and the energy density normalized
by $T^4$. For comparison, the Stefan-Boltzmann limit is also shown.
We can see the entropy density and the speed of sound on Fig. \ref{eos_sc}.
All quantities were directly obtained from the pressure as described 
previously (\ref{eq:escs}).

The light and strange quark number 
susceptibilities ($\chi_{ud}$ and $\chi_s$) are defined as
\begin{align}
\frac{\chi_{q}}{T^2}=\frac{1}{TV}\left.\frac{\partial^2 \log Z}{\partial \mu_{q} ^2
}\right|_{\mu_{q}=0}, 
\end{align}
where $\mu_{ud}$ and $\mu_s$ are the light and strange quark 
chemical potentials. With the help of the quark number operators 
one can split up the quark number susceptibilities into connected and
disconnected parts (see~\cite{AFKSZ} for details). The disconnected
parts turned out to be consistent with zero and negligible compared to
the connected parts. The connected part of $\chi_{ud}$ is twice the
isospin number susceptibility $\chi_{I}$.

\begin{figure}
\begin{center}
\includegraphics*[width=7.9cm]{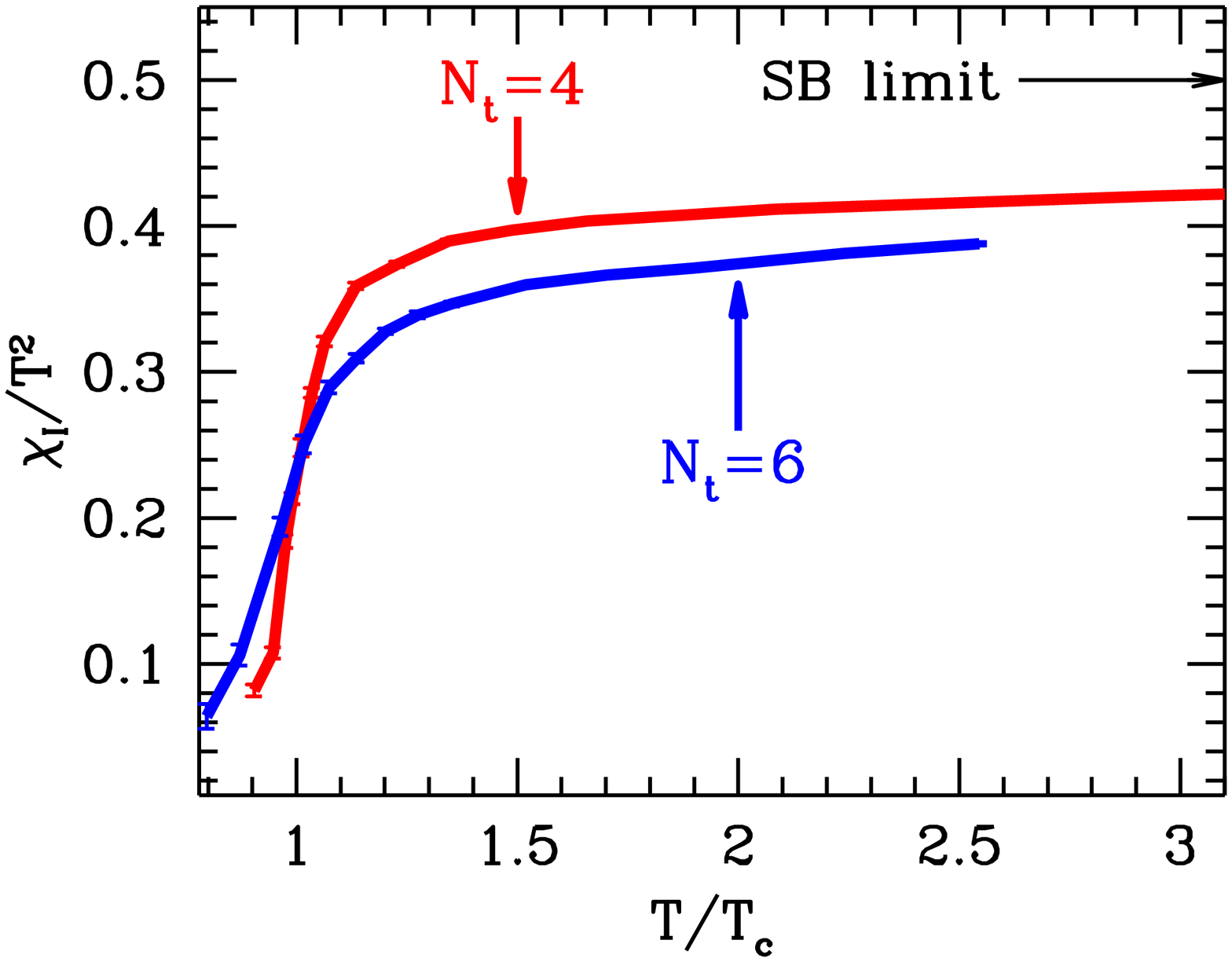}
\includegraphics*[width=7.9cm]{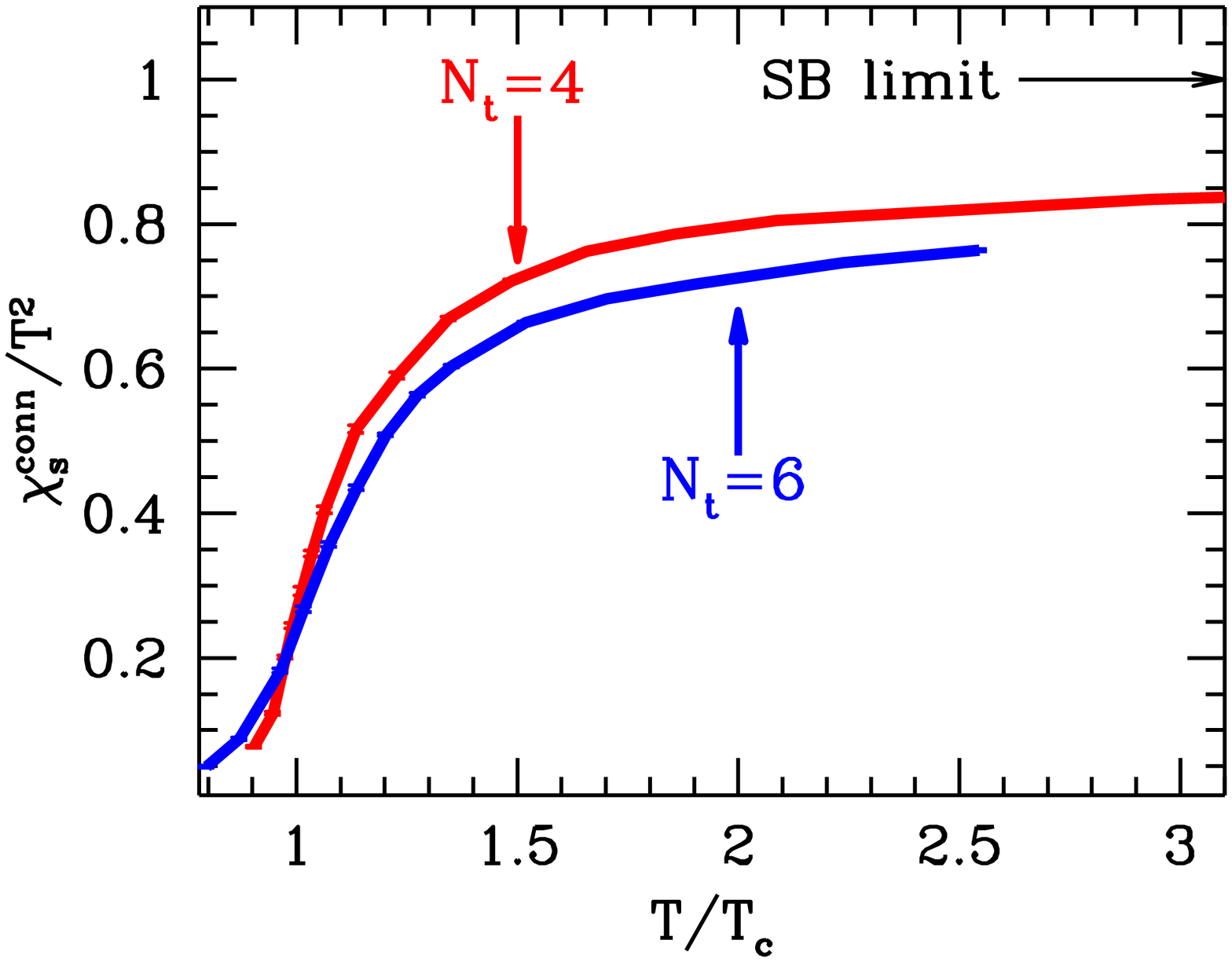}
\end{center}
\caption{\label{chi}
{\em Left:} the isospin susceptibility normalized by $T^2$.
{\em Right:} the connected part of the strangeness susceptibility. 
The labeling is the same as in Figure \ref{eos_pe}.
}
\end{figure}

Fig. \ref{chi} shows our results on $\chi_{I}$ and the connected
part of $\chi_s$. Again the data are scaled by $c_{cont}/c_{N_t}$.
The turning point of $\chi_I$ was used to define $T_c$.

\section{Summary}

The EoS can be determined using lattice QCD. While for the pure
gauge theory there are already continuum extrapolated results, 
in full QCD there are no final results yet.

Previous results using either staggered of Wilson fermions were
discussed. They suffer from several weaknesses.

In the second part of the paper new results on the EoS were presented.
Our analysis attempted to improve on previous
analyses by several means.
We used for the lightest hadronic degree of freedom the physical pion 
mass. We used two different sets of lattice spacings ($N_t$=4,6). 
The system was kept on the line of constant physics (LCP) instead
of changing the physics with the temperature.
Due to our smaller lattice spacing and particularly due to our 
stout-link improved fermionic action the
unphysical pion mass splitting was much smaller than in any 
previous staggered analysis. An exact calculation algorithm was applied.
 
We presented results for the pressure, energy density, entropy density,
speed of sound and the isospin and strangeness susceptibilities.

Although a continuum extrapolation could already be performed with the
current data, since the $N_t=4$ lattices are rather coarse (especially
around and below $T_c$) it would be safer if the EoS on even 
finer lattices ($N_t$=8) were obtained. Such an analysis would be a 
major step towards the final results for the equation of state.

\section{Acknowledgments} 
I thank Y. Aoki, Z. Fodor and K.K. Szab\'o for the nice collaboration 
and useful comments on the manuscript.
This work was partially supported by
OTKA Hungarian Science Grants No.\ T34980, T37615, M37071, T032501.
This research is part of the EU Integrated Infrastructure Initiative
Hadron physics project under contract number RII3-CT-20040506078.
The computations were carried out at E\"otv\"os University
on the 330 processor PC cluster of the Institute for Theoretical Physics
and the 1024 processor PC cluster of Wuppertal University.

\end{document}